\documentclass[conference]{IEEEtran}
\IEEEoverridecommandlockouts
\usepackage{cite}
\usepackage{amsmath,amssymb,amsfonts}
\usepackage{algorithmic}
\usepackage{graphicx}
\usepackage{textcomp}
\usepackage{xcolor}
\usepackage{todonotes}
\PassOptionsToPackage{hyphens}{url}
\usepackage{hyperref}
\usepackage{multirow}
\usepackage{comment}
\usepackage{ulem}
\usepackage{subcaption}
\usepackage{todonotes}

\def\BibTeX{{\rm B\kern-.05em{\sc i\kern-.025em b}\kern-.08em
    T\kern-.1667em\lower.7ex\hbox{E}\kern-.125emX}}

\begin{document}

\title{Utilization of Impedance Disparity Incurred from Switching Activities to Monitor and Characterize Firmware Activities}
%A Method for Monitoring and Categorizing Firmware Operations Utilizing the Impedance Disparity Incurred from  Switching Activities
%A method for monitoring and categorizing firmware operations utilizing the frequency characteristic of impedance disparity 

\author{\IEEEauthorblockN{Md Sadik Awal}
\IEEEauthorblockA{\textit{ECE Department} \\
\textit{Florida International University}\\
Miami, Florida, USA \\
mawal003@fiu.edu}
\and
\IEEEauthorblockN{Christopher Thompson}
\IEEEauthorblockA{\textit{ECE Department} \\
\textit{Weber State University}\\
Ogden, Utah, USA \\
cthompson2@mail.weber.edu}
\and
\IEEEauthorblockN{Md Tauhidur Rahman}
\IEEEauthorblockA{\textit{ECE Department} \\
\textit{Florida International University}\\
Miami, Florida, USA \\
mdtrahma@fiu.edu}
}

\maketitle

\begin{abstract}
The massive trend toward embedded systems introduces new security threats to prevent. Malicious firmware makes it easier to launch cyberattacks against embedded systems. Systems infected with malicious firmware maintain the appearance of normal firmware operation but execute undesirable activities, which is usually a security risk. Traditionally, cybercriminals use malicious firmware to develop possible back-doors for future attacks. Due to the restricted resources of embedded systems, it is difficult to thwart these attacks using the majority of contemporary standard security protocols. In addition, monitoring the firmware operations using existing side channels from outside the processing unit, such as electromagnetic radiation, necessitates a complicated hardware configuration and in-depth technical understanding. In this paper, we propose a physical side channel that is formed by detecting the overall impedance changes induced by the firmware actions of a central processing unit. To demonstrate how this side channel can be exploited for detecting firmware activities, we experimentally validate it using impedance measurements to distinguish between distinct firmware operations with an accuracy of greater than 90\%. These findings are the product of classifiers that are trained via machine learning. The implementation of our proposed methodology also leaves room for the use of hardware authentication.

\end{abstract}


\section{Introduction}
%This document is a model and instructions for \LaTeX.
%Please observe the conference page limits. 
%\todo{Our article focuses on firmware activities, but you have mentioned malware attacks. How are they connected? Please connect them. Alternatively or in parallel, focus on how our proposed approach is helping to detect malware attacks.}
Current estimates of the number of internet-connected devices speculate that there will be over 40 billion smart devices in 2025, with more than 30 billion of those being Internet of Things (IoT) devices \cite{iot_analytics_2021}. This is a major increase from the total and distribution of devices in 2015—of the 13.3 billion internet-connected devices in 2015, only 3.6 billion were considered IoT.
The underlying reason for this drastic increase in devices is attributed to the popularization of personal smart devices, home assistants, and the development of low-power wide-area networks. With this surge in linked devices, the possibility of cyberattacks increases. 
One of the most severe threats to IoT devices is a firmware attack. Recent work \cite{schmidt2016secure}, which encompasses the attack surfaces found in IoT devices, describes how firmware-based attacks can be leveraged for control hijacking, reverse engineering to obtain sensitive data, eavesdropping on sensitive packets, and creating system vulnerabilities to insert malware. Even if the firmware cannot be reverse-engineered, firmware updates can be exploited to distribute malware \cite{abdul2018comprehensive}.

%Recently, AT\&T's Alien Labs issued a report on the BotenaGo, a variant of the Mirai malware \cite{van2017techniques}. These attacks aim to establish backdoors in IoT devices and routers for eventual integration into a botnet \cite{caspi_2021}. In a new research, the Forescout Cybersecurity team investigates R4IoT, a novel technique for gaining early access to IoT devices that exploits IoT devices. These compromised devices target other networked devices for malware distribution \cite{forescout_security_group_2022}.

%\textcolor{red}{Introduce firmware; add a paragraph}
Firmware is a form of embedded software that supports fundamental device functionalities \cite{firmware_def1}. To begin with, when a device is turned on, the firmware is the first one to run and send the necessary instructions for the device to communicate with other devices or function properly. Without it, even the most basic of devices will be rendered inoperable. To prevent this from happening, the firmware is typically stored on an Erasable Programmable Read Only Memory (EPROM) or flash memory chip. But with technological advancement, firmware becomes obsolete even before the hardware does, as it needs to be updated to improve security, add new features, address issues, and support new protocols and standards. It is possible to exploit this flexibility, however, to allow malicious firmware updates \cite{firmware_flash_eprom}.

Malicious firmware and hardware components create an unacceptable security channel on embedded systems. According to \cite{caspi_2021}, firmware patches are capable of thwarting potential threats. This solution is appropriate only when the manufacturer is aware of the problem. It would be great if the system could detect and respond to threats on its own. However, embedded systems typically lack the same anti-malware protection measures as a standard personal computer. Since the embedded central processing unit (CPU) only has a limited number of resources, it is difficult to apply the majority of the standard security strategies. %\todo{What are the security strategies?}
For instance, the authors of \cite{goodUSB} modify the operating system to prevent USB-based attacks with direct human supervision, in which a USB stick is used to inject malicious code into secured device firmware in order to covertly compromise the system's confidentiality and reliability. The underlying cause of the issue is the implicit assumption that all extraterrestrial hardware is inherently trustworthy. Another work in \cite{SBAP} uses a remote software-based technique to validate the integrity of the peripherals linked to the system.
Similarly, in the paper \cite{SWATT}, software-based attestation is introduced to authenticate the software of a device. Here, the complete contents of the memory are checked using a checksum function. Therefore, a less resource-intensive solution that can monitor the firmware's performance in order to detect any anomalies is necessary.

\begin{comment}
\todo{Mention some existing side-channel analysis along with their limitations.} Existing side-channel analysis (SCA) varies from the method proposed in this study for detecting firmware activities in embedded systems by requiring a less sophisticated measuring setup and minimal external noise0 %\todo{If you make such claim, you need to justify it by comparing your proposed approach with other existing approaches.}. This method decreases the amount of equipment as well as the amount of labor that is necessary to detect malware on embedded devices. 
For instance, power SCA and electromagnetic SCA require specialized equipment and modifications to the circuitry \cite{PowerSCA}, \cite{EMSCA}. It is also important to have the necessary skills in order to deduce the relevant information from the data acquired by SCA. %Whether our approach reduces complexity by directly connecting to the MCU to analyze the side channel with minimal circuit modification.
\end{comment}

In this paper, we present a side-channel technique for observing % \todo{What are we detecting? Choose a different word.}
the activities performed by an embedded system. This is accomplished by measuring the effective impedance of the  microcontroller unit (MCU) across a range of frequencies while it is operating. As the MCU executes different instructions, the effective impedance of the MCU varies. This is the foundation of our strategy for detecting various firmware operations. After that, we classify the type of firmware that is operating on the system by utilizing several machine learning methods.

\begin{comment}
In this paper, we present a side-channel technique for detecting the activities performed by an embedded system. \textbf{The following are our contributions to this paper:
\begin{itemize}
    \item We postulate that the switching activity due to firmware changes the overall impedance between the source and ground. 
    \item We utilize the reflection coefficients measured by the help of a vector network analyzer (VNA) to find out the response difference due to the shift in impedance of a microcontroller unit (MCU).
    \item We use an algorithm to find out only the minimum number of frequency points for the VNA to look at for the best accuracy.   
    \item We classify the type of firmware that is operating on the system by utilizing several machine learning methods.
\end{itemize}
}
\end{comment}

\section{Related Works}
%\subsection{Background}
%\subsection{Related Works}
Previous works discuss and show the feasibility of detecting firmware activities through side channel analysis. The authors of \cite{wilt_baker_papadakis_2020} demonstrate a work to classify the operating system, distinguish which software was running and differentiate different malware types running on a single board computer through EM emissions and an RF probe with near perfect accuracy.  
Similarly, authors of \cite{pham_marion_mastio_heuser_2021} propose a method for identifying malware and classifying it while it is executing on a Raspberry Pi. Their method uses comparable methodologies and achieves results that are comparable to those obtained by the authors of \cite{wilt_baker_papadakis_2020}. 
The authors of the paper \cite{vedros_makrakis_kolias_xian_barbara_rieger_2021} detect single instruction malicious code injections in the firmware of an Arduino Mega with collected EM emissions. This collected data was passed through a k-Nearest Neighbor model, where they were able to classify single-instruction injections.  
In \cite{agrawal_chen_hollingsworth_hung_izmailov_koshy_liberti_mesterharm_morman_panagos_etal_2018}, a framework for detecting anomalous code executions using a combination of machine learning and statistical training techniques is presented.

%\textcolor{red}{In contrast, we use a vector network analyzer (VNA) to monitor and categorize firmware operations. The VNA-based side-channel information has been used to detect malicious modifications of hardware  \cite{Zhu_Shan_etal:2022,8605668}.} 
The impedance-based side-channel information has been used to detect malicious modifications of hardware   \cite{Zhu_Shan_etal:2022,8605668,nearfield,DasEtAl:2019,narasimhan2010multiple}.
The authors of \cite{Zhu_Shan_etal:2022} use a vector network analyzer (VNA) to detect hardware Trojans, recycled printed circuit boards (PCBs), malicious components,  and counterfeit processors. The underlying method to detect hardware anomalies on a board relies on measuring the equivalent impedance from various locations on the board. Their process of finding hardware anomalies used the Fr\'echet between a standard circuit frequency and an anomalous frequency response. The authors of \cite{8605668} implement a hardware modification detection %\todo{Hardware detection?} 
method by modeling the circuit as a resistor-capacitor circuit. The basis for the anomaly detection method relies on supplying the circuit with an AC voltage and measuring the current. Capacitance measurements are used to detect malicious hardware modification by comparing the modified device's reactance to a standard. 
Authors of \cite{nearfield} detect and classify certain types of physical tampering events employing a wide-band antenna as a nearfield probe. The concept is founded on the premise that changes in the radio channel affect the frequency characteristics of an antenna. The work is backed by both a theoretical foundation and empirical validation.
The authors in \cite{DasEtAl:2019} explore the frequency region in which these complementary metal-oxide semiconductor (CMOS) gates emit unintentional side channel information. As current travels through the various metallic traces in the chip, the tiny distances between these metal etchings cause electromagnetic radiation to be unintentionally emitted, which suggests a range of frequencies to be monitored to detect the firmware running on the embedded system. 
The side channel-based analysis of \cite{narasimhan2010multiple} establishes a relationship between active mode current and maximum operational frequency. The authors show that this relationship can be used to detect hardware Trojans.

%\sout{Thus far, we have reviewed some research that uses a variety of methodologies for identifying firmware and hardware level activities and anomalies.} 
In this article, we will present our proposal, which is based on the impedance response of the processing unit and may be used to identify and classify different sorts of firmware activities.

\section{Theoretical Analysis} %\todo{Instead of using "Proposed Methodology", choose a different title that represents this Section or the article.}
In this section, we discuss the theory that underpins our proposed method for evaluating the viability of detecting %aberrant 
firmware activities using a VNA. %By detecting the different effective impedance produced by the switching of a microcontroller, the VNA can be used to identify malicious activities.
We propose that this can be accomplished by identifying the change in effective impedance caused by the switching activity arising from the firmware operations of a microcontroller. 

%\todo{Introduce firmware and write statements on how this firmware affects hardware operations and switching activities. All programs run on hardware.}
The firmware of a device is a specialized type of computer software that offers low-level control for the device's specific hardware in an embedded system. It contains instructions to assist the hardware start-up, communicate with other devices, and execute basic input/output functions.
The instructions are executed on the system's hardware, which contains thousands, if not millions, of logic circuits. By switching these logic circuits between their on and off states, instructions are executed.

\begin{figure*}[htbp]
\centering
\begin{subfigure}[b]{0.3\linewidth}
	\centering
  	\includegraphics[scale=0.36]{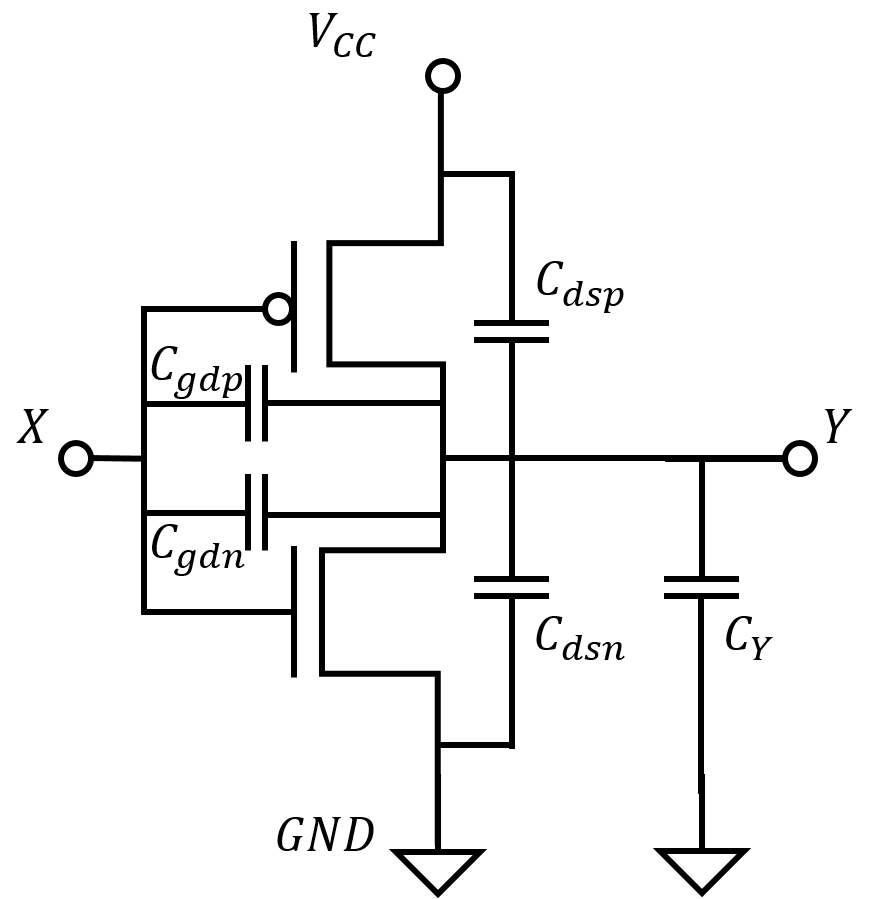}
  	\caption{}
  	\label{fig:inv_equiv_C}
\end{subfigure}
\begin{subfigure}[b]{0.3\linewidth}
	\centering
  	\includegraphics[scale=0.36]{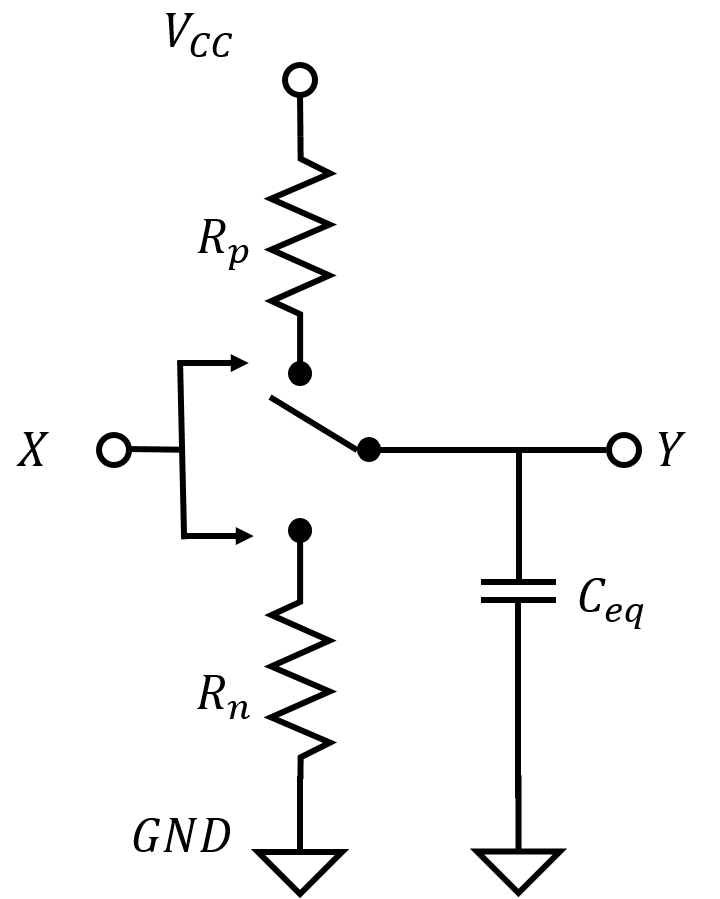}
  	\caption{}
  	\label{fig:inv_equiv_R}
\end{subfigure}
\begin{subfigure}[b]{0.3\linewidth}
	\centering
  	\includegraphics[scale=0.36]{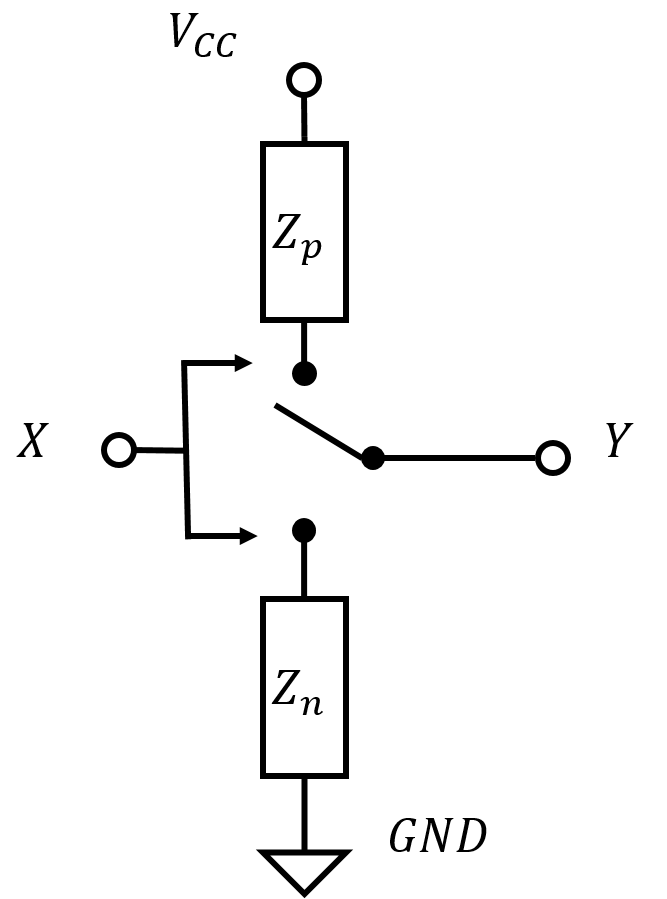}
  	\caption{}
  	\label{fig:inv_equiv_Z}
\end{subfigure}
\caption{ CMOS inverter (a) with the parasitic capacitances, (b) switch model of dynamic behavior, (c) equivalent impedance circuit.}
\label{fig:inv_equiv2}
\end{figure*}

%Many thousands or even millions of logic gate circuits are used in the fabrication of digital circuits such as microcontrollers and microprocessors. 
 %\todo{If you address previous comment, I am not sure the next sentence is necessary. Furthermore, a single sentence has two "fact"s.} %The fact that they have become so common in computer architecture is largely attributable to the fact that they consume very little power while they are in a stationary condition.

The architecture of the logic gates in integrated chips relies on CMOS circuits.
%\todo{Mention at some point that CMOS inverter or NOT gate is a universal logic gate and one of the building blocks of a digital circuit.} 
CMOS inverter/NOT gate, one of the basic universal logic gates, is a fundamental building block in digital logic circuits.
Digital logic circuits compute the tasks by regulating the metal–oxide–semiconductor field-effect transistors (MOSFETs) of the CMOS circuits. These MOSFETs are activated by applying a voltage to the gate, which regulates the drain current. A thin silicon oxide layer isolates the gate of a MOSFET from the drain and source, and by inverting the substrate between the drain and source, a parasitic diode is formed. In addition to exhibiting resistance, MOSFETs are characterized by the presence of capacitance between their terminals. This property is a direct consequence of the structure of MOSFETs. The capacitance of the gate oxide film determines the capacitance of the gate to drain and gate to source terminals. Parasitic diode junction capacitance determines drain to source capacitance.

\begin{comment}
\begin{figure}[htbp]
\centering
\begin{subfigure}[b]{.45\linewidth}
    \centering
	\includegraphics[scale=0.33]{Figures/inv_equiv.png}
	\caption{}
	\label{fig:inv_equiv}
\end{subfigure}%
\begin{subfigure}[b]{.45\linewidth}
	\centering
  	\includegraphics[scale=0.33]{Figures/inv_equiv_C.png}
  	\caption{}
  	\label{fig:inv_equiv_C}
\end{subfigure}
\caption{ (a) CMOS inverter, (b) with the parasitic capacitances}
\label{fig:inv_equiv1}
\end{figure}
\end{comment}

%In order to illustrate the proposed methodology, \todo{This sentence does not provide a complete message.} 
Fig.~\ref{fig:inv_equiv_C} depicts an equivalent circuit of CMOS containing parasitic capacitances. In this example, $C_{gdp/n}$ represents the gate to drain capacitance, $C_{dbp/n}$ represents the drain to bulk parasitic capacitance (the diffusion capacitance) for the PMOS/NMOS, and $C_Y$ represents the capacitance at the output wire \cite{DigitalIntegratedCircuits}, \cite{sedra_smith_2004_digital}. The expressions of the capacitance are presented in Table~\ref{tab:cap_eq}. The parameters associated with Table~\ref{tab:cap_eq} and their definitions are presented in Table~\ref{tab:cap_def}.

The output $Y$ of the CMOS inverter varies based on the input $X$ during the steady state. As shown in Fig.~\ref{fig:inv_equiv_R}, $R_{p}$ and $R_{n}$ are the effective on-resistance values for the PMOS and NMOS, respectively. Here, $C_{eq}$ represents the output capacitance, which is the aggregate of all parasitic capacitance in the CMOS inverter. Hence, $C_{eq} = \sum (C_{gdp}, C_{gdn}, C_{dbp}, C_{dbp})$. Let $R_{lin,p/n}$ and $R_{sat,p/n}$ represent the effective on-resistance in the linear and saturation regions, respectively, for the PMOS/NMOS, \cite{FPGABasedSystemDesign}, where, 
\begin{align}
    R_{lin,p/n} &= \frac{ \frac{1}{2} (V_{D} - V_{S} - V_{t,p/n}) } {\frac{3}{8} k'_{p/n} (\frac{W}{L})_{p/n} (V_{D} - V_{S} - V_{t,p/n})^{2}} \label{eq:R_lin} \\
    R_{sat,p/n} &= \frac{V_{D} - V_{S}}{\frac{1}{2} k'_{p/n} (\frac{W}{L})_{p/n} (V_{D} - V_{S} - V_{t,p/n})^{2} } \label{eq:R_sat}
\end{align}
Here, $k'_{p/n}$ represents the trans-conductance, $(W/L)_{p/n}$ represents the aspect ratio, and $V_{t,p/n}$ represents the threshold voltage of the PMOS/NMOS. The voltage at the drain is denoted by $V_{D}$, while the voltage at the source is denoted by $V_{S}$. Using \eqref{eq:R_lin} and \eqref{eq:R_sat}, $R_{p}$ and $R_{n}$ can be estimated as follows:
\begin{align}
    R_{p/n} = \frac{1}{2} (R_{lin,p/n} + R_{sat,p/n}) \label{eq:R_p/n}
\end{align}
Each of the equivalent impedances, $Z_{p}$ and $Z_{n}$ in Fig.~\ref{fig:inv_equiv_Z}, consists of equivalent resistance $R_{p/n}$ and equivalent reactance $X_{p/n}$. Therefore, $Z_{p/n} = R_{p/n} + j X_{p/n} $, where $R_{p/n}$ is the resistance and $X_{p/n}$ is the reactance of the PMOS/NMOS. The combined parasitic capacitance of the PMOS/NMOS dominates the equivalent reactance $X_{p/n}$ \cite{DigitalIntegratedCircuits}. Therefore, $X_{p/n} \approx \frac{-1}{\omega C_{eq,p/n}}$, where $C_{eq,p/n}$ is the total equivalent capacitance of the PMOS/NMOS between the output node $Y$ and the ground node $GND$.

\begin{table}[htbp]
\renewcommand{\arraystretch}{2}
\caption{Expression of the parasitic capacitance}
\label{tab:cap_eq}
\centering
\begin{tabular}{|c||c|} \hline
\bfseries Capacitor & Expression\\ \hline\hline
$C_{gdp}$ & $2C_{op}W_{p}$ \\ \hline
$C_{gdn}$ & $2C_{on}W_{n}$ \\ \hline
$C_{dsp}$ & $K_{bpp} AD_{p} CJ_{p} + K_{swp} PD_{p} CJSW_{p}$ \\ \hline
$C_{dsn}$ & $K_{bpn} AD_{n} CJ_{n} + K_{swn} PD_{n} CJSW_{n}$ \\ \hline
$C_{Y}$ & From Extraction\\ \hline
\end{tabular}
\end{table}

\begin{table}[htbp]
\renewcommand{\arraystretch}{2}
\caption{Definitions of the parameters of parasitic capacitance}
\label{tab:cap_def}
\centering
\begin{tabular}{|c||c|} \hline
\bfseries Parameter Symbol & Definition \\ \hline\hline
$C_{op}, C_{on}$ & Overlap capacitance per unit width \\
\hline
$W_{p/n}$ & Width of PMOS/NMOS \\ \hline
$K_{bpp/n}$ & Capacitor linearization factor of bottom plate \\ \hline
$AD_{p/n}$ & Area of drain \\ \hline
$CJ_{p/n}$ & Bottom junction capacitance  \\ \hline
$K_{swp/n}$ & Capacitor linearization factor of sidewall \\ \hline
$PD_{p/n}$ & Perimeter of drain  \\ \hline
$CJSW_{p/n}$ & Sidewall junction capacitance \\ \hline
\end{tabular}
\end{table}

The output terminal $Y$ is connected to either the impedance $Z_p$ or $Z_n$ depending on the gate input $X$. 
Due to process variation and MOSFET geometry, the impedance of the MOSFETs differs from each other. As a consequence, the CMOS inputs govern the impedance measured between the source voltage node $V_{CC}$ and the ground voltage node $GND$. Other logic circuits in the CPU exhibit similar input/output impedance shifting characteristics due to switching as well.

As the CPU executes its instruction set, inputs are applied to the gate terminals of logic gates in order to perform operations. In response to changes in the inputs, the impedance between the source and ground nodes of logic gates alters. Therefore, it is to be expected that the overall impedance that exists between the source node and the ground node would shift in response to the various operations carried out by the CPU. In other words, the switching activity of the CMOS results in a change in the impedance between the source and ground.

%Based on that statement \todo{What specific statement?}, 
\textbf{To summarize,  }
we conclude %\todo{theorize or conclude} 
that as the CPU changes the output states of the logic gates, the size of the overall effective impedance between the source and ground node should change depending on the task it is performing. Therefore, distinct impedance signatures should be generated by distinct types of firmware activities.

\section{Experimental Setup}
%This section describes the experimental setup, data processing and the findings using ML models.

The goal of the experimental set-up is to evaluate the feasibility of the technique that we have proposed for monitoring the activities of a CPU. The Arduino Due \footnote{The Arduino Due \cite{arduino} is powered by the AT91SAM3X8E microcontroller with 96 KB of SRAM and 3.3V of operational voltage. It operates at 84 MHz and has a total DC output current of 130 mA across all of its I/O lines}, a VNA \cite{pocketvna_2021}, and a breadboard with some LEDs and resistors to limit the current drain through the LEDs form up the infrastructure that is used in the study to collect data. A 5V power supply is used to power the Arduino Due. Afterwards, a capacitor %with a rating of 250 pF 
is used with the probe to connect the VNA to the 3.3V pin of the microcontroller so that the observation can be recorded. Since the Arduino Due operates on 3.3V, monitoring the virtual impedance of the CPU while the firmware is being executed can be performed by sampling the 3.3V pin.

\begin{figure}[htbp]
    \centerline{\includegraphics[width = 0.46\textwidth]{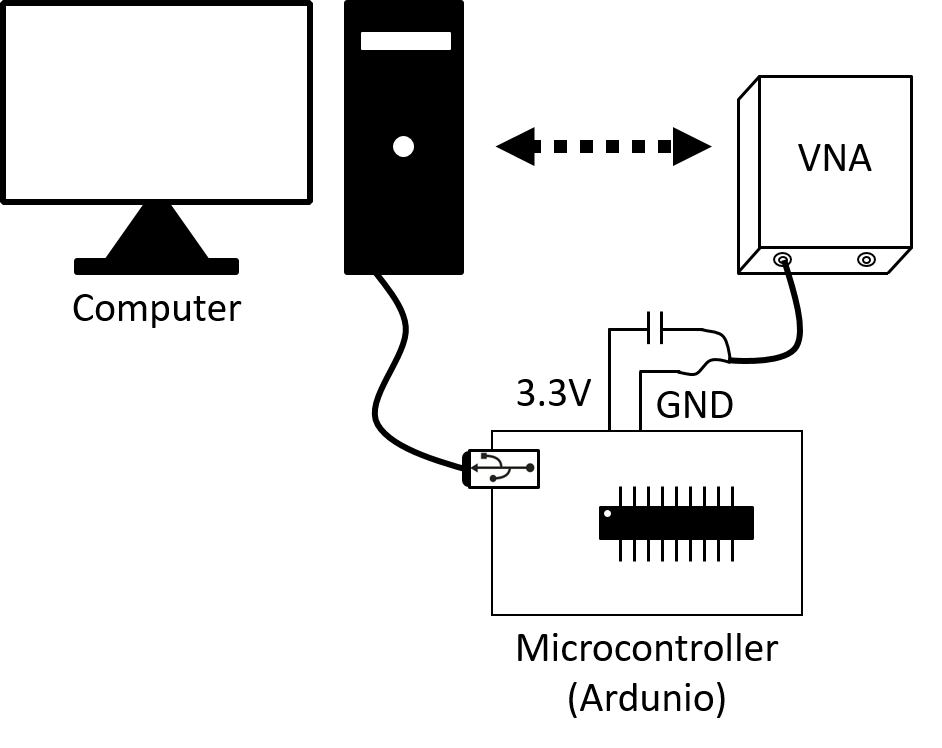}}
    %\caption{Experimental setup for collecting data. }
    \caption{Data collection from the microcontroller with a VNA. }
    \label{fig:layout}
\end{figure}

The VNA has a measuring range of 500 kHz to 4 GHz. We use the VNA to measure the forward reflection coefficient, a parameter that characterizes the amount of reflected wave in the transmission medium, to calculate the impedance at 10,000 linearly divided frequency points throughout its spectral region. To eliminate the DC component from measured signals, we put a capacitor in series. This data collected by the VNA is subsequently processed for classification analysis.
Fig.~\ref{fig:layout} depicts a diagram that describes the %connections of the 
experimental setup. The VNA is used to perform the measurements. A computer stores the measured signals and uploads different firmware to the Arduino microcontroller.

%\todo{change the fig. title from ``Diagram of the experimental setup." to something else.}
For the purpose of developing firmware operations, we compose C/C++ code \footnote{The firmware activity codes for this experiment are accessible at \url{https://github.com/ChristopherThompsonUT/ArduinoRepo}} in the Arduino IDE and upload it to the MCU.
This experiment considers the following four possible firmware operations:

\begin{enumerate}
  \item Case 1: In this instance, the MCU is powered on but no operation is being performed.
  \item Case 2: Three resistors and LEDs are connected to the digital pins of the Arduino and switched on and off at a rate of 1 kHz. These pins, as well as eight other digital I/O pins, are connected with additional resistors to ground. The purpose of this is to imitate an operation for the microcontroller that draws the total maximum allowable I/O current from all of the I/O pins (130 mA).
  \item Case 3: In this instance, the MCU executes two distinct programs. The first program periodically turns an LED on and off at 1 Hz. In the background, as the MCU performs this operation, another program exponentiates random four-digit numbers. These numerals are raised to powers of one through three hundred. This latter program replicates possibly dangerous firmware operating in the background of the system, and is similar to the device acting as a botnet host. Essentially, the system preserves the illusion of regular operation but executes unwanted activities. 
  This is akin to the system being compromised by the Mirai or BotenaGo malware \cite{van2017techniques}. The BigNumber Arduino Library \cite{BigNumber} is used to reliably store and handle 256-bit values in order to manipulate these numbers.
  \item Case 4: The Advanced Encryption Standard (AES) is utilized to replicate the situation in which a MCU is executing a sensitive task. The Arduino Cryptography Library \cite{ArduinoCryptography} is used to encrypt and decrypt strings of ten ASCII characters that are generated at random.
\end{enumerate} 
Using these four scripts, we collect a total of 445 observations per case, for a grand total of 1,780 observations. %In the collected dataset, each observation has 10,000 values that represent the forward reflection coefficients at each of the 10,000 linearly divided frequency points spanning from 500 kHz to 4 GHz. 
Each observation contains the forward reflection coefficients at the most relevant and dominating frequency points extending from 500 kHz to 4 GHz (frequency range of the VNA) in the collected dataset. The step-by-step process for locating those frequency locations is described in greater detail in \ref{sec:scanning}. These frequency points help us to reduce the amount of redundant information in the dataset.

Let $\tau$ represent the reflection coefficient, $Z_{r}$ represent the reference impedance, and $Z_{\tau}$ represent the impedance obtained from the reflection coefficient of the medium. Eq.~\eqref{eq:impedance} describes the relation between $Z_{\tau}$ and $\tau$.
\begin{equation}
    Z_{\tau} = Z_{r} \left(\frac{1+\tau}{1-\tau}\right)
    \label{eq:impedance}
\end{equation}
To compute the impedance, we use the recorded reflection coefficients and \eqref{eq:impedance}.
The reference impedance, $Z_r$, for the VNA in this case is 50 $\Omega$. We employ impedance measurements generated from reflection coefficients in the classification stage.

\section{Result and Analysis}
\begin{figure}[htbp]
    \centerline{\includegraphics[width = 0.5\textwidth]{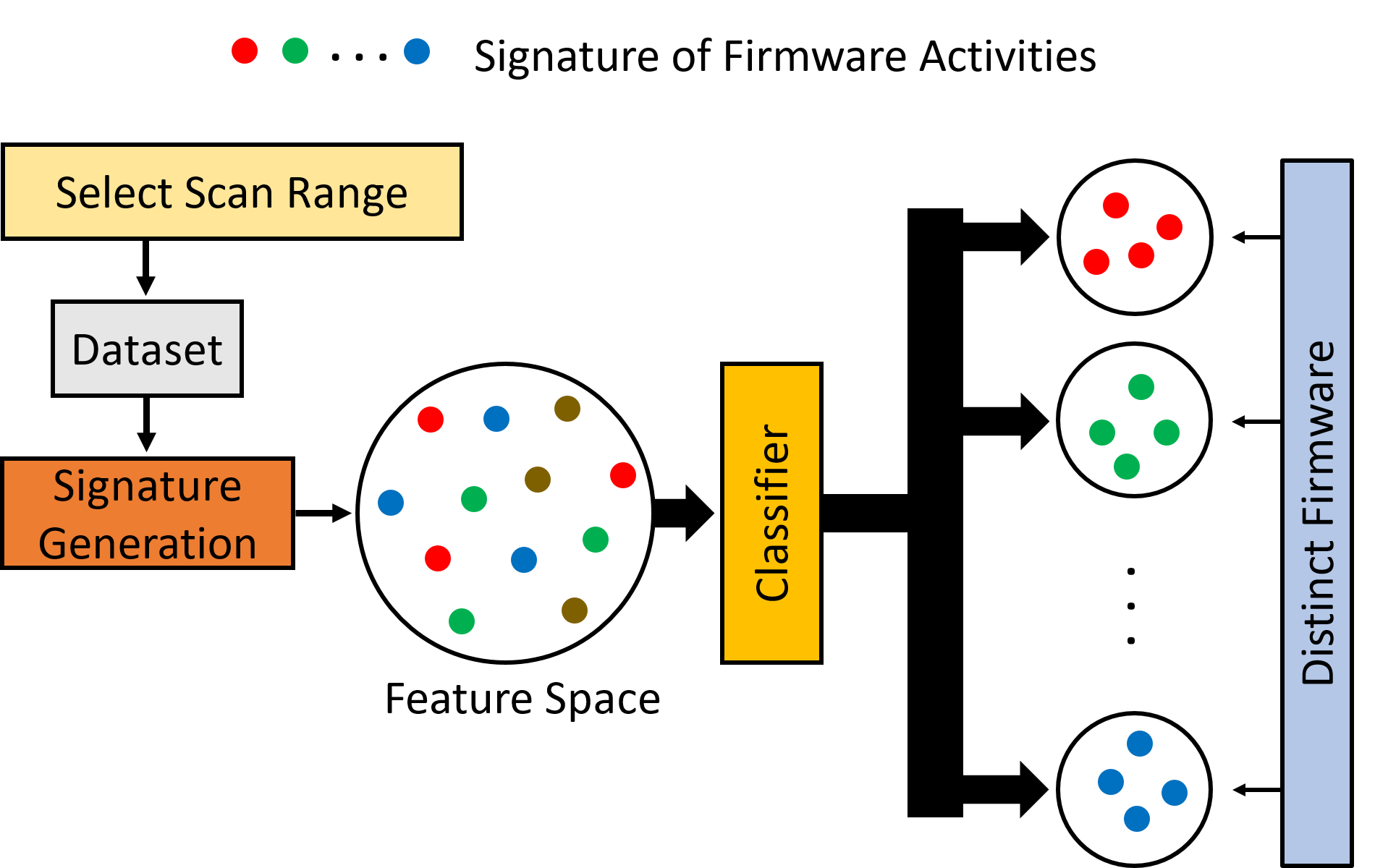}}
    \caption{Steps in firmware activity detection and classification}
    \label{fig:flow}
\end{figure}

This section begins by describing our analysis approach for detecting firmware activity from a dataset. Following this is an overview of what we studied. Fig.~\ref{fig:flow} illustrates the categorization process in a simplified form. In order to make use of any machine learning model, an appropriate dataset is required that is composed of the fingerprints of the activities carried out by the firmware. The steps involved in this process are outlined in the subsequent subsections.
\subsection{Selecting Scanning Region}
\label{sec:scanning}
\begin{figure}[htbp]
    \centerline{\includegraphics[scale=0.43]{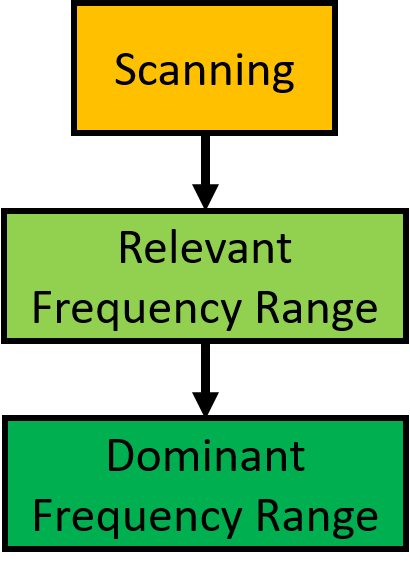}}
    \caption{Selection of appropriate frequency points}
    \label{fig:freq_sel}
\end{figure}

The VNA in our experimental setup scans 10,000 frequency points that are linearly split and range from 500 kHz to 4 GHz in frequency. This section focuses on selecting the appropriate frequencies to investigate for signatures generated from various firmware activities. We desire to find the optimal frequency points in order to reduce the computation time and increase the number of relevant sample points. The procedure is depicted graphically in Fig.~\ref{fig:freq_sel}.

To investigate the frequency points, we collect the responses of 10,000 of them while performing various tasks on the microcontroller. A subset of about 20\% of the total frequency points is then investigated to determine which ones are most strongly connected to various firmware operations. This allows us to determine the frequencies that are relevant. In order to do this, we compute the Pearson correlation coefficients between the responses of each frequency point and specific firmware activity. On the basis of the correlation coefficients, the first 1968 frequency points (about 20\% of all frequencies points) that are highly correlated with firmware activities are selected. These 1968 frequency points also reflect the 70\% correlation coefficients for the correlation coefficient with the highest value. After that, we select the maximum number of frequencies that are linearly independent of one another in order to evaluate the dominant frequency points. This allows us to reduce the number by an additional factor. In this step, we once again utilize the Pearson correlation coefficients between the frequencies. These subsequently lowered frequency groups exhibit correlation values of less than 90\%. Following this two-step process of frequency selection, we are able to identify 338 frequency points %(3.38\%) 
that adequately describe the activities of the firmware without significantly compromising accuracy. 
These 338 frequency points represent only a 3.38\% of the frequency range of the VNA.
During the data collection process with the VNA, we concentrate on only these specific frequency points.
%method 3 = 0.70, 0.90	34 features	1968, 338

\subsection{Feature Selection and Classification}
%We describe the feature selection and machine learning classifiers 
We train and test the classifier models with the help of the MATLAB Statistics and Machine Learning Toolbox. 
As discussed in the preceding subsection, we construct two distinct datasets, one for the training and the other one for the testing, utilizing the 338 frequency points that are the most significant and dominant in explaining the actions of the firmware. One of these datasets is used for training, and the other for testing.

As depicted in Fig.~\ref{fig:flow}, we first use the training dataset to produce signatures from various firmware actions in order to deduce any meaningful use from the dataset. In order to get this collection of signatures prepared for the machine learning (ML) classifier, we run a principal component analysis (PCA) \cite{jackson2005user} on the training data to determine which features are accountable for 95\% of the variance in the data. This allows us to prepare the signatures for the ML classifier. To reduce training time and improve the classifier's accuracy, it is desirable to select a subset of the available features. PCA facilitates the generation of a reduced collection of features without compromising the variance of the data. Using PCA, we can explain at least 95\% of the variation in the training dataset using only 34 principal components. In the training phase of the machine learning classifier, these 34 principal components serve as features. To prevent overtraining and to stabilize the models, classifiers are trained using five-fold cross-validation. %We use five fold cross validation for the training phase. 
For the purposes of validating the models, we use the testing dataset, which comprises 30\% of the total collected data. 

\begin{table*}[htbp]
\renewcommand{\arraystretch}{1.75}
\caption{Comparison of accuracy and precision metric}
\label{tab:result_table}
\centering
\begin{tabular}{|c||c|c|c|c|c|c|}
\hline
\bfseries Classifier & F1 Score (Train) & F1 Score (Test) & Precision & Recall & Specificity & Accuracy \\ %macroAVG
\hline\hline
 SVM (Kernel: Gaussian) \cite{SVMGaussian}              & 92.4\% & 93.7\% & 93.9\% & 93.6\% & 97.9\% & 93.6\% \\ \hline
 SVM (Kernel: Cubic) \cite{SVM_Gaussian_Cubic}             & 92.1\% & 93.3\% & 93.7\% & 93.2\% & 97.7\% & 93.3\% \\ \hline
 SVM (Kernel: Quadratic) \cite{SVM_Quad}   & 92.4\% & 91.8\% & 92.2\% & 91.7\% & 97.2\% & 91.8\% \\ \hline
 Quadratic Discriminant \cite{QuadDis} & 92.1\% & 91.5\% & 91.7\% & 91.4\% & 97.1\% & 91.4\% \\ \hline
 Subspace KNN \cite{SubKNN}           & 91\%   & 90.9\% & 91.2\% & 90.8\% & 96.9\% & 90.8\% \\ \hline
\end{tabular}
\end{table*}
To evaluate the performance of the ML classifiers, we calculate the precision, recall, specificity, accuracy and F1 scores \cite{powers2020evaluation,witten2005practical}. Let $TP_{i}$, $TN_{i}$, $FP_{i}$, and $FN_{i}$ be the true positives, true negatives, false positives, and false negatives predicted by a given classifier for prediction class $i$, respectively. 
In this context, a true positive is the number of correctly predicted positive values, whereas a false positive is the number of inaccurately predicted positive values. Additionally, a false negative is the number of values incorrectly predicted as negative, whereas a true negative is the number of correctly predicted negative values. 
Precision measures the proportion of positive class predictions that correspond to the actual positive class. Thus, precision evaluates the accuracy for the minority class and is calculated as \cite{powers2020evaluation}, 
\begin{equation}
Precision = \frac{1}{i} \sum_{i} \left(\frac{TP_{i}}{TP_{i} + FP_{i}}\right) \label{eq:precision} 
\end{equation}
Recall quantifies the number of correct class predictions generated for all positive examples in the training set \cite{powers2020evaluation}. The recall indicates that positive forecasts were missed. 
\begin{equation}
    Recall = \frac{1}{i} \sum_{i} \left(\frac{TP_{i}}{TP_{i} + FN_{i}}\right) \label{eq:recall} 
\end{equation}
The specificity of a classifier is the ratio between the amount of data that is accurately classified as negative and the actual amount of data that is negative \cite{powers2020evaluation}. 
\begin{equation}
    Specificity = \frac{1}{i} \sum_{i} \left(\frac{TN_{i}}{TN_{i} + FP_{i}} \right)\label{eq:specificity}
\end{equation}
The accuracy of a dataset's predictions is the ratio of correct predictions to the total number of predictions. The accuracy increases as the number of correct predictions made by the system increases \cite{witten2005practical}.
\begin{equation}
    Accuracy = \frac{1}{i} \sum_{i} \left( \frac{TP_{i} + TN_{i}}{TP_{i} + TN_{i} + FP_{i} + FN_{i}}\right) \label{eq:accuracy}
\end{equation}
The F1 score provides a single value that addresses both precision and recall concerns in a single number. This is the mean harmonic of the two fractions \cite{witten2005practical}. A classifier's performance improves as its F1 score increases. 
\begin{equation}
    F1 = \frac{2 \times Precision \times Recall}{Precision+Recall} \label{eq:f1} 
\end{equation}
We use \eqref{eq:precision}-\eqref{eq:f1} to calculate the performance metrics and compare the performance of the classifiers. Table \ref{tab:result_table} illustrates the F1 score, precision, recall, specificity, and accuracy for five ML classifiers. %the accuracy of the five best-performing models. 
%The results of our investigation indicate that the models with the highest performance are the SVM Kernel, Quadratic SVM (QSVM), Cubic SVM, and Quadratic Discriminant. 
The classifiers perform very well and classify firmware activities with an F1 score, precision, recall, specificity, and accuracy of more than 90\%, indicating that our proposed method to detect firmware operations is statistically viable.

\section{Conclusion and Future Work}
In this study, we demonstrate that a VNA is capable of detecting aberrant activity in digital logic systems. This is accomplished by monitoring the effective impedance of an Arduino Due, measuring the values of the forward reflection coefficients using a VNA over a range of significant frequencies, extracting features from these, and passing these feature sets to various machine learning models. The top-performing classifier is able to determine which firmware is currently being executed by the system by making use of 34 features, achieving an accuracy of 93.2\% during training and 94.8\% during testing, respectively.

In the future, we plan to apply more feature extraction techniques to determine if this can potentially improve the classifiers' accuracy. %Multiple observations of a scenario can be averaged to remove undesired noise as a potential method for improving accuracy. 
% feature selection: SVM-REF, Genetic Algorithm. extraction: autoencoders, wavelet scattering. reverse engineering: and operation, xor operation..
In addition, we intend to determine if our proposed method may be utilized in reverse engineering to recover critical hardware and algorithmic data. Since we can discover firmware anomalies in this way, we can investigate hardware Trojans. Lastly, our proposed strategy may also investigate how injected components influence the computing processes of embedded systems.

\section*{Acknowledgment}

This work was supported partly by the National Science Foundation under Grant Nos. 2150248, 2214108, and 2114200.  %Any opinions, findings, and conclusions or recommendations expressed in this material are those of the author(s) and do not necessarily reflect the views of the National Science Foundation

\normalem
\bibliographystyle{IEEEtran}
\bibliography{IEEEabrv,references}
\end{document}